\documentclass[aps,prc,twocolumn,notitlepage,nofootinbib]{revtex4-1}

\usepackage{graphicx}
\usepackage[figuresright]{rotating}
\usepackage{makecell}
\usepackage{bm,amsmath,amssymb}
\usepackage[mathscr]{eucal}
\usepackage{multirow}
\usepackage{xcolor}
\usepackage{mathrsfs}
\usepackage{mathtools}
\usepackage{slashed}
\usepackage{graphics}
\usepackage{graphicx}
\usepackage{subfigure}
\usepackage{dsfont}
\usepackage{longtable}
\usepackage{bbm} 
\usepackage{float}
\usepackage{tcolorbox}
\usepackage{lipsum}
\usepackage{cancel}
\usepackage{enumerate}
\usepackage{diagbox}
\usepackage[breaklinks,colorlinks,citecolor=citcolor,urlcolor=blue,linkcolor=lcolor]{hyperref}
\definecolor{lcolor}{rgb}{0.5,0,0}
\definecolor{citcolor}{rgb}{0,0.3,0.0}

\newcommand{\pT}{p_{\rm T}}

\newcommand{\Tcf}{T_{\rm CF}}
\newcommand{\tfs}{\tau_{\rm fs}}

\newcommand{\ie}{\textit{i.e.}}
\newcommand{\eg}{\textit{e.g.}}
\newcommand{\raa}{R_{\rm AA}}

\newcommand{\ccb}{c\bar{c}}
\newcommand{\rmd}{{\rm{d}}}

\newcommand{\Ds}{\mathcal{D}_s}
\newcommand{\NN}{\mathrm{NN}}

\begin{document}

\title{Nonperturbative Heavy-Flavor Transport Approach for Hot QCD Matter}

\author{Tharun Krishna}
\affiliation{Department of Physics and Astronomy and Cyclotron Institute, Texas A\&M University, College Station, TX 77843-3366, U.S.A}

\author{Yu Fu}
\affiliation{Department of Physics, Duke University, Durham, NC  27708-0305, U.S.A}

\author{Weiyao Ke}
\affiliation{Key Laboratory of Quark and Lepton Physics (MOE) \& Institute of Particle Physics, Central China Normal University, Wuhan 430079, China}

\author{Steffen A. Bass}
\affiliation{Department of Physics, Duke University, Durham, NC  27708-0305, U.S.A}

\author{Ralf Rapp}
\affiliation{Department of Physics and Astronomy and Cyclotron Institute, Texas A\&M University, College Station, TX 77843-3366, U.S.A}

\date{\today}

\begin{abstract}
The heavy charm and bottom quarks are unique probes of the transport properties of the quark-gluon plasma (QGP) and its hadronization in high-energy nuclear collisions. 
A key challenge in this context is to embed the interactions of the heavy quarks in the expanding medium compatible with the strong-coupling nature of the QGP, and thus to unravel the underlying microscopic mechanisms. In the present work we progress toward this goal by combining recent $T$-matrix interactions for elastic scattering with an effective transport implementation of gluon radiation, and apply these in a Langevin framework in a viscous hydrodynamic evolution. Hadronization of heavy quarks is evaluated using a modern recombination model with 4-momentum conservation, supplemented with fragmentation constrained by data in proton-proton collisions. Deploying this approach to charm-hadron observables in Pb-Pb collisions at the LHC yields fair agreement with experiment while also identifying areas of further systematic improvement of the simulations and its current input.      
\end{abstract}

\maketitle

%%%%%%%%%%%%%%%%%%%%
{\it Introduction}.
%%%%%%%%%%%%%%%%%%%%
The investigation of the transport properties of hot Quantum Chromodynamics (QCD) matter as formed in ultra-relativistic heavy-ion collisions (URHICs) is at the forefront of contemporary research in nuclear physics. Key phenomenological evidence from hydrodynamic simulations of the bulk medium evolution~\cite{Heinz:2013th,Shuryak:2014zxa,Bernhard:2019bmu} and from heavy-flavor (HF) diffusion calculations~\cite{Rapp:2018qla,ALICE:2021rxa,He:2022ywp} indicates that pertinent transport parameters, \ie, ratio of the shear viscosity to entropy density ($\eta/s$) and of the HF diffusion coefficient to the thermal wavelength, $\Ds (2\pi T)$  ($T$: temperature), are close to lower limits conjectured from the strong-coupling limit of quantum field theory~\cite{Danielewicz:1984ww,Kovtun:2004de}.  
A fundamental objective is to connect these findings to the underlying QCD forces in quark-gluon plasma (QGP). This requires the use of nonperturbative interactions in a framework that rigorously implements quantum effects as the latter are expected to be critical in the vicinity of the strong-coupling limit.
To realize this objective in the context of HF spectra in URHICs
requires a quantitative framework for HF transport and hadronization. Such a framework is generally believed to consist of three (or five) main components~\cite{Rapp:2018qla}: heavy-quark (HQ) diffusion in the QGP, HQ hadronization, and a reliable space-time simulation of the fireball (pre-equilibrium evolution and hadronic rescattering are expected to be less important due to a short time duration and small interaction rates, respectively). 

The theoretical efforts toward a comprehensive approach have made tangible progress in recent years~\cite{Rapp:2018qla,Zhao:2023nrz} leading, \eg, to rather quantitative constraints on the HQ diffusion coefficient from combined model comparisons to the nuclear modification factor and elliptic flow of $D$-mesons~\cite{ALICE:2021rxa,CMS:2017vhp}. While the hadronization mechanism plays an important role in the simultaneous description of these observables, more direct constraints have been obtained by measuring additional charm hadrons, in particular $D_s$ mesons and $\Lambda_c$ baryons~\cite{ALICE:2021bib,ALICE:2021kfc}. 
Significant advances in experimental precision and breadth of HF observables are now enabling stringent tests of model descriptions, both on the phenomenological side and the underlying theory that ultimately determines the momentum and $T$-dependencies of the HQ interactions.

This letter reports on recent advances by combining components from two approaches while also improving on individual components. Specifically, we synthesize microscopic HQ $T$-matrix interactions in a strongly coupled QGP (sQGP)~\cite{Liu:2017qah,Tang:2023tkm} plus the resonance recombination model~\cite{Ravagli:2008rt} with the Trento plus viscous-fluid dynamics approach for bulk evolution~\cite{Shen:2014vra,Bernhard:2019bmu} and HF transport that merges elastic diffusion and radiative interactions~\cite{Ke:2018tsh}. For the first time we employ HQ $T$-matrix interactions with an underlying potential constrained by recent lattice-QCD (lQCD) 
data~\cite{Altenkort:2023eav,Altenkort:2023eav} (rather than using the internal energy~\cite{He:2019vgs}), thereby also reproducing the equation of state from lQCD as used in the hydrodynamic evolution~\cite{HotQCD:2014kol}. 
%\WK{ABout radiation and dead cone:} 
In addition, we refine the effective transport implementation of radiation with an improved interference behavior as found for static 
media~\cite{Ke:2018jem}.

%%%%%%%%%%%%%%%%%%%%%%%%%%%%%%%%%%%
{\em Langevin Approach in sQGP}.
%%%%%%%%%%%%%%%%%%%%%%%%%%%%%%%%%%%%
In the LIDO model~\cite{Ke:2018tsh} the time evolution of the HQ phase space distribution, $f_Q(t,x,p)$, can be formally expressed as
\begin{align}
\frac{\rmd f_Q}{\rmd t} = \mathcal{D}[f_Q] + C_{1\leftrightarrow 2}[f_Q]  \ , 
%+  C_{2\leftrightarrow 2}[f_Q] + C_{2\leftrightarrow 3}[f_Q] \ .
\end{align}
%For interactions with a small change in HQ momentum $q< \Qcut$, a Fokker-Planck approach is adopted 
with a diffusion operator given by
% \begin{align}
% \mathcal{D}=  -\frac{\partial}{\partial p_i}(A_i - \frac{1}{2}\frac{\partial}{\partial p_j}B_{ij})   \ , 
% \end{align}
\begin{align}
\mathcal{D}=  \frac{\partial}{\partial p_i}\Big(A(p) p_i + \frac{\partial}{\partial p^i}B(p)\Big)   \ . 
\end{align}
The elastic HQ transport coefficients (friction and momentum diffusion) figuring in this operator are computed from heavy-light $T$-matrices within the quantum many-body theory developed in Ref.~\cite{Liu:2017qah}. In particular, they encode off-shell properties of the medium through the use of light-parton spectral functions that are self-consistently calculated, while the bare masses in the underlying Hamiltonian are adjusted to reproduce the lQCD equation of state within the thermodynamically conserving quantum many-body formalism of Luttinger-Ward and Baym~\cite{Luttinger:1960ua,Baym:1961zz}. Schematically, the underlying system of Dyson-Schwinger equations can be written as
\begin{eqnarray}
T_{ij} &=& V_{ij} + \int V_{ij} G_i G_j T_{ij} \\
G_i &=& 1/(p_0 - \varepsilon_i(p) - \Sigma_i(p_0,p) )  \\
\Sigma_i &=& \int \sum\limits_m T_{im} G_m f^m \ , 
\label{eq:Tmat}
\end{eqnarray}
where the parton indices $i$ and $j$ denote both heavy and light anti-/quarks or gluons; the summation over $m$ in the selfenergy, $\Sigma_i$, is over thermal partons in the heat bath with corresponding distribution functions, $f^m$ (Fermi or Bose). Since the single-particle propagators, $G_i$, depend on the selfenergies and the latter on the $T$-matrices, one has a selfconsistency problem in both heavy and light sectors that is solved by numerical iteration.

\begin{figure*}[t]
    \centering
    \includegraphics[width=0.49\linewidth]{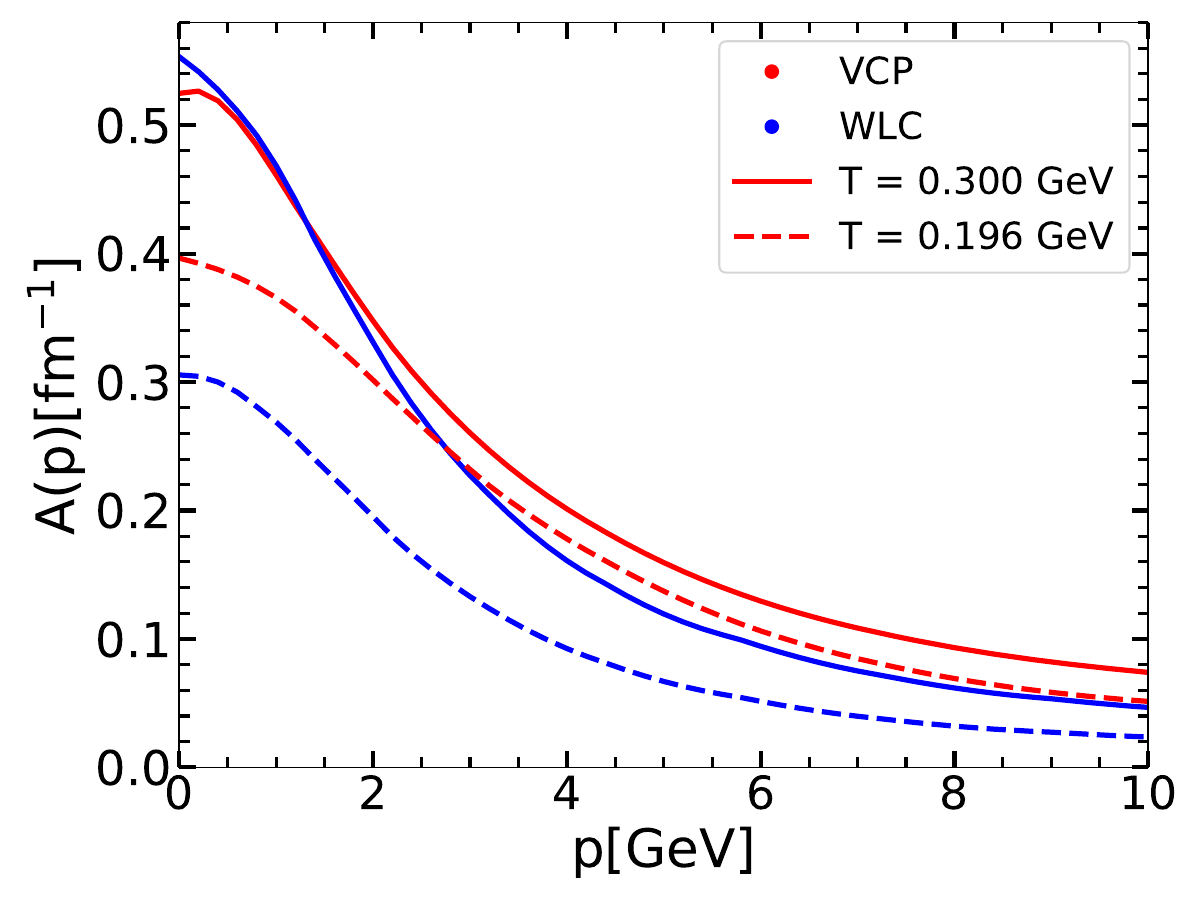}
    \includegraphics[width=0.49\linewidth]{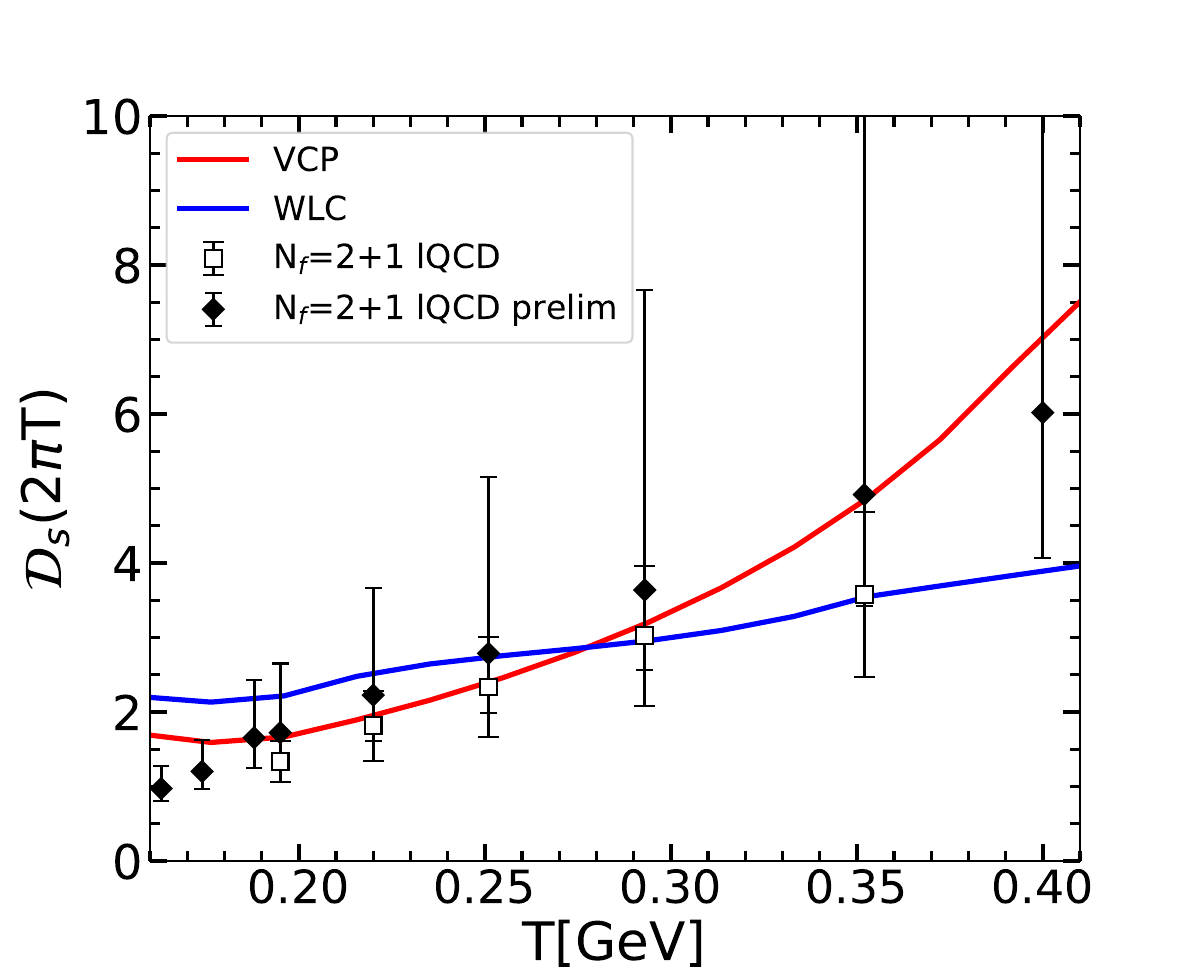}  
    \caption{Charm-quark friction (left panel) and spatial-diffusion coefficient (right panel) from selfconsistent heavy-light $T$-matrices in a strongly coupled QGP for the VCP (red lines) and WLC (blue lines) scenarios, compared to lQCD data~\cite{Altenkort:2023eav,Altenkort:2023oms,HotQCD:2025fbd}.}
    \label{fig:coeff}
   % \vspace{-0.05cm}
\end{figure*}

For the input potential (which is universal to both heavy- and light-parton interactions and includes relativistic corrections) we use two versions that are based on two sets of constraints from lattice QCD: (a) HQ free energies with a rather strong vector component in the confining potential deduced from spin and spin-orbit splittings in the vacuum quarkonium spectra (referred to as VCP), and (b) recent Wilson line correlators (referred to as WLC) which turn out to be particularly sensitive to the collisional widths of the heavy quarks in the QGP~\cite{Tang:2023tkm}, but prefer a somewhat weaker momentum dependence. These features are reflected in the resulting friction coefficient, $A(p)$, displayed in the left panel of Fig.~\ref{fig:coeff}. The predicted diffusion coefficient,
$\Ds=\frac{T}{MA(p=0)}$, shown in the right panel, agrees somewhat better with lQCD data for the VCP scenario~\cite{Altenkort:2023oms,HotQCD:2025fbd}. {The contribution from gluon radiation is negligible~\cite{Liu:2020dlt}.}
%At the smaller temperature, where both potentials are still rather close to the vacuum form, the larger interaction strength due to the larger vector component in the string force leads to an overall larger thermalization rate  in the VCP scenario, while at the higher temperature this effect is largely compensated by the weaker screening with temperature in the WLC scenario.
%We further display in Fig.~\ref{fig:Ds} the predicted values of the charm-quark spatial diffusion coefficient, which follows from the zero-momentum limit of the friction coefficient as $\Ds=T/MA(p=0)$.  In comparison to recent lattice-QCD results~\cite{Altenkort:2023oms,Altenkort:2023eav}, the VCP constraints yield better agreement at low temperatures, while the WLC constraints appear to give a better account of the rather weak temperature dependence suggested by the lattice data.
%
% \begin{figure*}[t]
%     \centering
%     \includegraphics[width=0.49\linewidth]{fig/Aofp.pdf}
%     \includegraphics[width=0.49\linewidth]{fig/spatial_diffusion_coefficient.pdf}  
%     %
%     \caption{Charm-quark friction (left panel) and spatial-diffusion coefficient (right panel) from selfconsistent heavy-light $T$-matrices in a strongly coupled QGP for the VCP (red lines) and WLC (blue lines) scenarios. }
%     \label{fig:coeff}
% \end{figure*}

Following previous studies (and supported by case studies with the Boltzmann equation~\cite{Rapp:2018qla}), we enforce the Einstein relation by setting the momentum diffusion coefficient (also referred to as $\kappa$) from the friction coefficient via $B=TE_pA(p)$, where $E_p=(M^2+p^2)^{1/2}$ is the HQ on-shell energy. The diffusing quark is treated as a quasiparticle, but the transport coefficients encode the full off-shell dynamics of the medium through the use of the broad thermal-parton spectral functions underlying the description of the equation of state (EoS)~\cite{Liu:2018syc}. This is critical for accessing the interaction strength of the broad heavy-light $D$-meson bound states that form below the nominal charm-light threshold as the temperature drops toward the hadronization transition.

%he thermodynamic $T$-matrix formalism; this includes, in particular, no constraints on momentum transfer in the pertinent scattering equation~\ref{eq:Tmat}.  In line with this picture, we retain in the LIDO model only the Fokker-Planck component governed by $\mathcal{D}[f_Q]$ and the corresponding diffusion-induced radiation term $\mathcal{C}_{1\leftrightarrow 2}[f_Q]$, while turning off the hard scattering and its associated radiation processes. T
%For gluon radiations from a light quark due to incoherent small-angle scatterings, 
The diffusion-induced radiation term, $\mathcal{C}_{1\leftrightarrow2}$, is simulated with the  collision rate method, where the rate for a $1\to2$ processes is given by
\begin{align}
\label{eq:diff-rate1to2-BH}
    R_{q\to qg} = \frac{dN^q_h}{dt dx d^2k_\perp}=  \frac{\alpha_s(k_{\perp}^2)C_F}{2\pi^2}  \frac{\hat{q}_S P_{gq}(x)}{(k_{\perp}^2 +m_{\infty}^2)^2} \ .
\end{align}
Here,  $P_{gq}(x)$ denotes the QCD splitting function and $m_{\infty}=m_D/\sqrt{2}$ the thermal gluon mass, regulating the collinear divergence. The in-medium running coupling,  
\begin{align}
\alpha_s(Q^2) = \frac{4\pi}{\beta_0}\Bigg{/} {\ln\!\left[ \frac{\max\{\,Q^2,(\mu \pi T)^2\,\}}{\Lambda_{\mathrm{QCD}}^2} \right]},
\end{align}
utilizes $\beta_0$=11$-\frac{2}{3}n_f$ with $n_f$=3, $\Lambda_{\mathrm{QCD}}$=0.2\,{GeV}, and a dimensionless parameter, $\mu$=1.5, controlling the characteristic medium scale.
%With the assumption that the running scale should be larger than the medium scale $\mu\pi T$
The jet transport coefficient is taken from the momentum diffusion coefficient as $\hat{q}_S$=$2\kappa$.

The expression in eq.~(\ref{eq:diff-rate1to2-BH}) applies to the Bethe-Heitler (BH) regime where the gluon formation time, $\tau_f$, is short compared to its mean-free-path, $\lambda$. In the Landau-Pomeranchuk-Migdal (LPM) regime where $\tau_f\gg \lambda$, we account for coherent scatterings with the medium. In perturbative-QCD effective kinetic theory~\cite{Arnold:2002ja,Arnold:2002zm}, the branching rate in the deep-LPM regime is reduced by a factor $\sim$ $\lambda/\tau_f$ compared to the BH limit. This led the authors of Ref.~\cite{Ke:2018jem} to introduce a prescription in the transport treatment where, subsequent to a 
$1\to2$ branching at time $t_0$,
%at time $t_0$ is initiated at a given, but 
the daughter partons are not instantly considered as independent, but the parton system continues elastic scatterings with the medium until a time $t$, where
\begin{align}
    t-t_0 = \tau_f\equiv \frac{2x(1-x)E}{k_{\perp}^2(t,t_0)+x^2M^2} \ ,
\label{eq:tauf}
\end{align}
%and the $x^2M^2$ term is only present for heavy quark gluon radiation. 
and $k_{\perp}^2(t,t_0)$ is the accumulated transverse-momentum broadening of the system due to elastic collisions during that period.
At time $t$, the physical branching is carried out with an acceptance proportional to $\lambda/\tau_f\sqrt{\ln(\tau_f/\lambda)}$, which suppresses emissions with long formation times, thus reflecting the destructive-interference LPM effect. Once accepted, the partons continue their evolution as independent quasi-particles.
For short formation times, $\lambda/\tau_f>1$, the acceptance probability saturates at unity, recovering the BH rate.

%s another consequence of the massive quark propagator, 
The LIDO model implements the HQ dead cone effect via a suppression factor, $\left(1+\theta_g^2/\theta_M^2\right)^{-n}$, relative to the gluon radiation rate of a light quark ($\theta_g$: gluon emission angle, $\theta_M = M/E_p$:  dead cone angle). The power exponent $n$=2 was employed in earlier work~\cite{Ke:2018jem}, but recent studies~\cite{Fu:2025} show that $n$=1 provides better consistency with calculations in soft-collinear effective theory~\cite{Kang:2016ofv}.

%thus providing a smooth interpolation between the latter and the LPM regimes.
%, and it was shown to agree well with theoretical calculations in the deep-LPM region for induced radiations in an infinite static medium, cf.~Ref.\cite{Ke:2018jem} for further details of this method.

Finally, following earlier studies \cite{Ke:2020clc}, we neglect 2$\to$1 gluon fusion processes: they are suppressed due to the dead cone effect at low HQ momentum while at high momentum the approach toward equilibrium renders them subdominant.
%energy loss effects from $1\to 2$ gluon splitting dominant over $2\to 1$. 
Further scrutiny of this approximation will be reported in forthcoming work~\cite{Fu:2025}.

%\RR{Following earlier studies, we neglect the 2$\to$1 back reactions: the deadcone effect suppresses them at low HQ momentum while at high momentum the approach toward equilibrium renders energy loss effects dominant. Further scrutiny of this approximation will be reported in forthcoming work~\cite{Fu:2025}.}

%%%%%%%%%%%%%%%%%%%%%%%%%%
{\em Heavy-quark hadronization}. 
%%%%%%%%%%%%%%%%%%%%%%%%%%
%After diffusion through partonic matter,  color-neutralize into hadrons. 
Due to a large ambient thermal-parton density, recombination processes play an important role in HQ hadronization at low and intermediate momentum. This has been supported early on by HF data in Au-Au(200\,GeV) collisions at RHIC,  especially for the elliptic flow~\cite{PHENIX:2006iih}. The recombination of heavy quarks with thermal light quarks enables the HF hadrons to directly inherit the $v_2$ of the bulk medium around the hadronization transition, where it is believed to be near its maximal value. 
Here, we focus on the resonance recombination model (RRM)~\cite{Ravagli:2007xx}, which possesses several effectual features. Being derived from an underlying Boltzmann equation, it conserves 4-momentum in the parton-to-hadron conversion and yields the correct equilibrium limit of the produced hadrons, also for expanding media with radial and elliptic flow~\cite{He:2011qa}. Its key ingredients are resonant cross sections for heavy-light parton scattering into $D$-mesons and charmed baryons, which naturally follow from the $T$-matrix interactions used in the diffusion part described above. It thereby enables a straightforward inclusion of excited states, which are likely important for the hadro-chemistry of the HF hadrons in nuclear collisions.
The RRM is supplemented by fragmentation processes which take over at 
high transverse momentum ($\pT$) and in $pp$ collisions.

The RRM equation for the phase space distribution (PSD) of a meson $M$ %from HQ-antiquark recombination 
can be obtained as~\cite{Ravagli:2007xx}
\begin{align}
f_M(x,\vec{p}_M) &= \frac{\gamma_M}{\Gamma_M} \int \frac{d^3\vec{p}_1}{(2\pi)^3} \frac{d^3\vec{p}_2}{(2\pi)^3} f_q(x,p_1) f_{\bar{q}}(x,p_2) 
\nonumber 
\\ 
&\quad \times \sigma(s) v_{\text{rel}}  (2\pi)^3 \delta^3(\vec{p}_M - \vec{p}_1 - \vec{p}_2),
\label{eq:rrm1}
\end{align}
where the resonance cross section, $\sigma(s)$, for $q + \bar{q} \to M$ is of relativistic Breit-Wigner form 
with a meson width $\Gamma_M$; $f_{q,\bar{q}}$ are the thermal-quark/antiquark PSDs, $v_{\text{rel}}$ is the relative velocity, and $\gamma_M = E_M / m_M$. We implement the RRM on a hydrodynamic hypersurface of the LIDO model at a hadronization temperature of $T_H=160$\,MeV, employing an event-by-event processing of the $c$ quarks from the diffusion calculation that retains the correlations between their momentum and spatial point of hadronization~\cite{He:2019vgs}. These extend the $\pT$ reach of the RRM component significantly and affect both open-charm~\cite{He:2019vgs} and charmonium~\cite{He:2022ywp} spectra for $\pT\gtrsim 5$\,GeV.
%RRM reproduces equilibrium Cooper–Frye spectra when using thermal quark inputs. For off-equilibrium charm quarks, it serves as a linear-response approximation. The invariant meson spectrum is given by
% \begin{align}\label{Cooper-frye-RRM}
% \frac{d N_M}{p_T dp_T d\phi dy} = \int \frac{p \cdot d\sigma}{(2\pi)^3} f_M(\vec{x}, \vec{p}),
% \end{align}
% where $d\sigma$ is the hypersurface element.

To evaluate the partition between recombination and fragmentation, a recombination probability, $P_{\text{rec}}(p^*_c)$, is determined selfconsistently from the RRM expression, eq.~(\ref{eq:rrm1}), for each $c$ quark from the diffusion simulation in the fluid rest frame with momentum $p^*_c$. 
%\begin{equation}\label{RRM Recombination probability}
%P_{\text{rec}}(p^*_c) = P_0 \int \frac{d^3 p_q}{(2\pi)^3} f_q(p_q) \frac{\gamma_M}{\Gamma_M} \sigma(s) v_{\text{rel}}.
%\end{equation}
After all charm-hadron states are summed up, an overall normalization is fixed to render this probability equal to one at $p_c^*$=0 (as there is no energy available for fragmentation).
The masses of the light ($q$=$u,d$), strange ($s$) and $c$ quarks are taken as $m_{q,s,c} = 0.3, 0.4, 1.5$\,GeV, and 
$\Gamma_M = 0.1$\,GeV. The RRM includes all charmed-hadron states listed by the particle data group (PDG)~\cite{ParticleDataGroup:2018ovx} plus charm-baryon states predicted by the relativistic quark model~\cite{Ebert:2011kk}, with branching ratios guided by the PDG.  
For $c$ quarks that do not recombine, fragmentation is carried out in the lab frame with a probability $1-P_{\text{rec}}(p^*_c)$ using fragmentation functions (FFs) from HQ effective theory (HQET).
%\begin{equation}
%\begin{aligned}
%\mathcal{D}_{Q \rightarrow H_P} &= N \frac{r z (1-z)^2}{[1 - (1 - r) z]^6} \Big[6 - 18(1 - 2r)z \\
%&\quad + (21 - 74r + 68r^2)z^2   \\
%&\quad -2(1 - r)(6 - 19r + 18r^2)z^3 \\
%&\quad +3(1-r)^2(1-2r+2r^2)z^4
%\Big] \ .
%\end{aligned}
%\end{equation}

\begin{figure}[!t]
    \centering
    \includegraphics[width=0.99\linewidth]{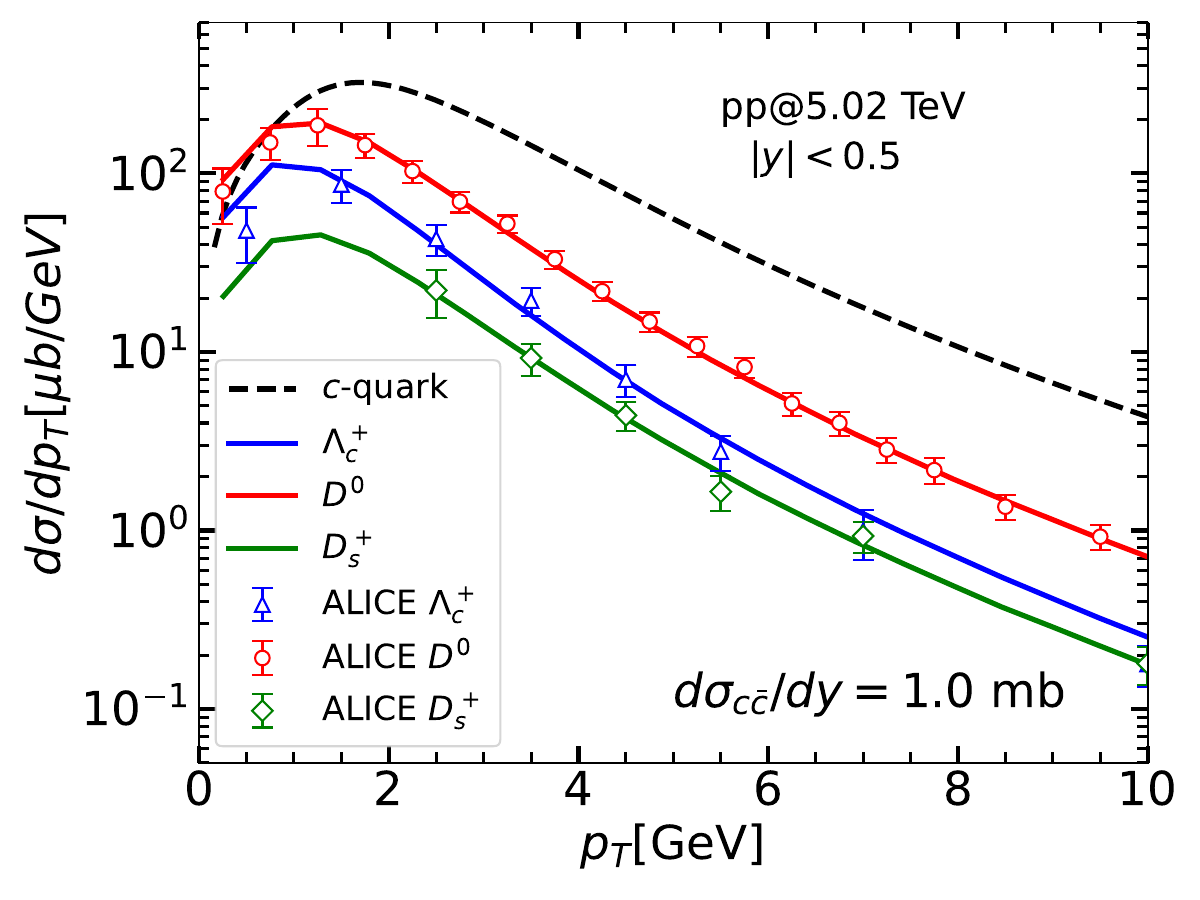}
    \caption{Production cross sections of prompt $D^0$ (red), $D_s^+$ (green) and $\Lambda_c^+$ (blue) hadrons in $pp$ collisions at $\sqrt{s_{\NN}}$=5.02\,TeV, using FONLL $c$-quark spectra (dashed line) and HQET fragmentation with statistical weights from the SHM, compared to ALICE data~\cite{ALICE:2019nxm,ALICE:2022exq}.}
    \label{fig:pp_fonll}
\end{figure}

The FFs are determined in fits to $\pT$ spectra of $D$, $D^*$, $D_s^+$ and $\Lambda_c$ hadrons in $pp$ collisions at the LHC (summarized in Fig.~\ref{fig:pp_fonll}) following Ref.~\cite{He:2019tik}. They are based on $c$-quark spectra from the fixed-order next-to-leading logarithm (FONLL) framework and carry
%are combined with HQET pertinent parameters quoted there.
%The $r$ parameters for the ground-state meson ($r_D$=0.1) and baryon ($r_{\Lambda_c}=0.16$) are tuned while for excited states a mass scaling is employed as $r_M/r_{D^0} = ((m_M-m_c)/m_M)/(m_{D^0}-m_c)/(m_{D^0})$ (and likewise for baryons). 
hadro-chemical weights for the individual charm-hadron species determined from a statistical hadronization model with $T_H=170$\, MeV. A single overall normalization, $N$, is applied, amounting to the total charm cross section, $d\sigma_{\ccb}/dy$. We also require a strangeness suppression factor of $\gamma_s$=0.7 to reproduce the yield of the charm-strange $D_s^+$ mesons in $pp$ collisions. This is slightly larger than the value of 0.6 in previous work~\cite{He:2019tik}, caused by revised branching ratios of excited $D_s^*$ whose masses are above the $DK$ threshold. Previously all these $D_s^*$'s were decayed into $D_s\pi$, while here, following available information from the PDG, we take a branching ratio of $\sim$80\% for $D_s^*\to DK$ above the $DK$ threshold. For the RRM component in URHICs, the thermal strange-quark PSDs in the QGP are evaluated with $\gamma_s=1$.
%in terms of the inclusive $D$-meson spectra and the $D_s/D$ ratio.

%After diffusion through the QGP, each $c$-quark’s fluid rest frame momentum is computed to obtain the recombination probability, $P_{\text{rec}}(p^*_c)$. This, in turn, gives a ``fragmentation probability" as $1-P_{\text{rec}}(p^*_c)$.
%, which partitions the recombination and fragmentation yields. 

%%%%%%%%%%%%%%%%%%%%%%%%%%%%%%%%%%%
{\it Bulk Evolution Model}.
%%%%%%%%%%%%%%%%%%%%%%%%%%%%%%%%%%%%
A realistic bulk medium evolution for URHICs is a prerequisite for a reliable simulation of HQ transport within, as it provides the local temperature and flow fields for the transport coefficients that govern HQ dynamics.
In the present work, we utilize a 2+1-dimensional viscous hydrodynamical evolution, which consists of several stages.
Initial conditions are generated at $\tau=0^+$ based on a partition of participant and binary-collision profiles~\cite{Moreland:2014oya}. 
The pre-equilibrium stage is approximated by free streaming until the start of hydrodynamics at $\tfs$~\cite{Liu:2015nwa} for the QGP evolution with an up-to-date lattice EoS~\cite{Shen:2014vra,HotQCD:2014kol}.
The hydrodynamic medium is converted into an ensemble of light hadrons via sampling a hypersurface of fixed temperature of $\Tcf=152$~MeV utilizing the Cooper-Frye formula. Note that this does not have to be the same as the temperature where the medium converts from partons to hadrons, which is not sharply defined. 
The hadronic ensemble is subject to further rescattering and decays utilizing the Ultra-Relativistic Quantum Molecular Dynamics (UrQMD) model~\cite{Bass:1998ca,Bleicher:1999xi}. 
All of these stages involve parameters that have previously been calibrated to reproduce a vast array of bulk observables at the LHC~\cite{Bernhard:2019bmu}, providing a description of the bulk evolution of the QGP at high precision.
We will, however, defer the study of the impact of both the pre-equilibrium and hadronic phase on the HF spectra~\cite{Fu:2025}, as their effect on the nuclear modification factor and elliptic flow is expected to small~\cite{Das:2024vac,Chesler:2013cqa,Mrowczynski:2017kso}. 
%For this particular study we use a fixed elastic rescattering cross section of 5 mbarn to provide an approximate matching to the heavy quark transport coefficient at $T_C$ \footnote{a more precise matching to the heavy quark transport coefficient with realistic hadronic rescattering cross sections is planned for a future publication}
%The bulk medium is initialized by the deposition of energy and matter density during the initial moments of the collision. In this study, the pre-hydrodynamic evolution of the medium is modeled using the TRENTO framework, a phenomenological model that provides the initial geometry and energy density distribution for ultra-relativistic heavy-ion collisions. The system then undergoes free-streaming for a proper time $\tau_{fs}$, which provides the initial condition for subsequent hydrodynamic evolution. Here, we employ the (2+1)-dimensional viscous hydrodynamics model (VISHNU) to describe the dynamical evolution of the thermalized bulk medium or QGP. As the hydro system reaches the critical temperature $T_c$, light hadrons are sampled from the corresponding hypersurface of the hydro. This particlization process can be described by the Cooper-Frye formula.

%%%%%%%%%%%%%%%%%%%%%%%%%%%%%%%%%%%%%%%
{\it Comparison to data}.
%%%%%%%%%%%%%%%%%%%%%%%%%%%%%%%%%%%%%%%%%
% \begin{figure*}[t]
%   \centering
%   \begin{minipage}[t]{0.5\textwidth}
%     \includegraphics[width=\linewidth]{fig/D0_RAA_30_50_updated.pdf}
%     %\caption*{(a) Caption for first figure}
%   \end{minipage}
%     \hspace{-0.8cm}
%   \begin{minipage}[t]{0.5\textwidth}
%     \centering
%     \includegraphics[width=\linewidth]{fig/D0_RAA_0_10_updated.pdf}
%     %\caption*{(b) Caption for second figure}
%   \end{minipage}
% \vspace{-0.05cm}
%     \begin{minipage}[t]{0.5\textwidth}
%     \centering
%     \includegraphics[width=\linewidth]{fig/D0_v2_30_50_updated.pdf}
%     %\caption*{(a) Caption for first figure}
%   \end{minipage}
%     \hspace{-0.8cm}
%   \begin{minipage}[t]{0.5\textwidth}
%     \includegraphics[width=\linewidth]{fig/D0_v2_0_10_updated.pdf}
%     %\caption*{(b) Caption for second figure}
%   \end{minipage}
%   \caption{Nuclear modification factor (upper panels) and elliptic flow (lower panels), $R_{AA}$, of $D^0$ mesons at mid rapidity in PbPb collisions at $\sqrt{s_{\NN}}=5.02\text{TeV}$, calculated using the Resonance Recombination model and the Heavy quark effective theory for recombination and fragmentation respectively. The red and blue curves are the calculations with the in-medium potentials in T-matrix formalism, VCP, and WLC, respectively. Experimental $\raa$ and $v_2$ data are from Refs.~\cite{ALICE:2021rxa} and \cite{ALICE:2020iug,CMS:2020bnz}, respectively.
%   }
%   \label{fig:PbPb-D0}
% \end{figure*}
% % 
\begin{figure*}[t]
\centering
\includegraphics[width=\linewidth]{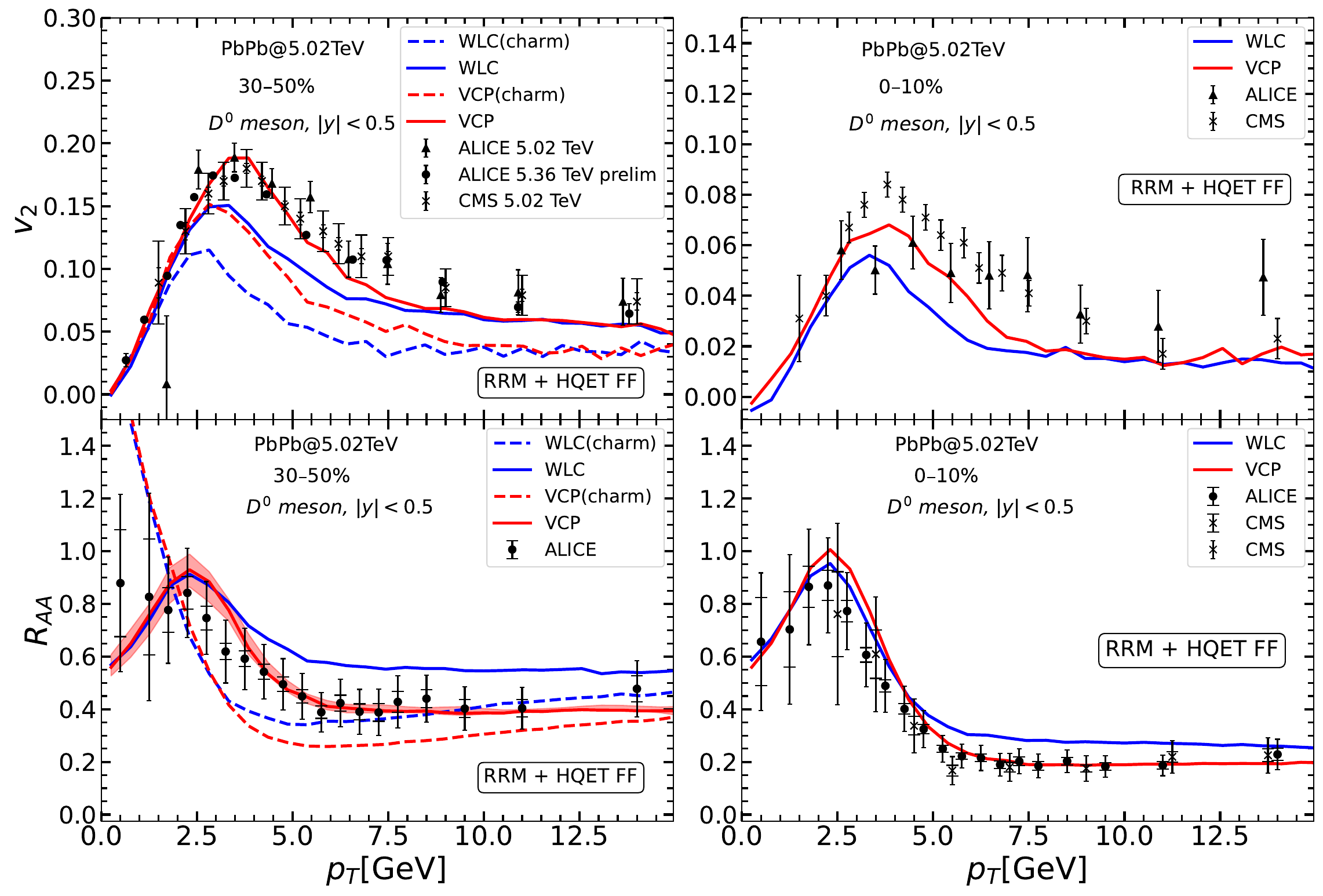}
\caption{Nuclear modification factor, $\raa$  (lower panels), and elliptic flow, $v_2$ (upper panels), of $D^0$ mesons at mid-rapidity in Pb-Pb collisions at $\sqrt{s_{\NN}}=5.02\text{ TeV}$, calculated from our integrated HF transport and hadronization approach using $T$-matrix interactions with VCP (red lines) or WLC (blue lines) constraints. The red band in the lower left panel illustrates a shadowing range of 70–80\% (the default is 75\%).  Experimental $\raa$ and $v_2$ data are from Refs.~\cite{ALICE:2021rxa,CMS:2017qjw} and \cite{ALICE:2020iug,CMS:2020bnz}, respectively. {The dashed lines in the left panels are for $c$-quarks just prior to hadronization.}}
\vspace{-0.05cm}
\label{fig:PbPb-D0}
\end{figure*}
We will focus our analysis on open HF meson production in Pb–Pb collisions at a center-of-mass energy of $\sqrt{s_{\mathrm{NN}}} = 5.02$ TeV.
The time evolution of the HQ distribution function in the QGP requires initial conditions in both position and momentum space. We assume that all $c$-quarks are produced upon initial impact of the incoming Pb nuclei with a Glauber model binary-collision profile for their spatial distribution. In momentum space, we employ the FONLL~\cite{Cacciari:2001td} spectra determined in $pp$ fits, augmented by nuclear shadowing with a $\pT$-dependence taken from Ref.~\cite{Emelyanov:1997guf} and an integrated suppression of 25\% in semi-/central Pb-Pb collisions, {compatible with recent measurements~\cite{ALICE:2022exq} (when varying the suppression over 20-30\%, the maximum in the $\raa$ varies by up to 5\%, while the $v_2$ is affected very little).}
At the hydro thermalization time, $\tfs$, we commence the QGP diffusion with elastic $T$-matrix interactions and radiation as described above, followed by the RRM-HQET hadronization model. As mentioned above, we neglect effects from the pre-equilibrium and hadronic phase in the present study.

From our HF hadron spectra, we compute the nuclear modification factor, $\raa$, and elliptic-flow coefficient, $v_2$. The former amounts to the ratio of the $\pT$ spectra in AA collisions to that in $pp$, normalized to the number of primordial $NN$ collisions, $N_{\text{coll}}$, at a given centrality. 
% \begin{equation}
% R_{AA}=\frac{1}{\langle N_{\text{coll}}\rangle} \frac{\rmd N_{\text{AA}}/\rmd p_T}{ \rmd N_{\text{pp}}/\rmd p_T} \ .
% \end{equation}
The $v_2$ follows from the Fourier decomposition of the azimuthal-angle dependence of the $\pT$ spectra, relative to the reaction plane (for simplicity, we have neglected event-by-event fluctuations in this study). 
%\begin{equation}
%E_H\frac{\rmd^3 N}{\rmd^3 p}=\frac{1}{2\pi}\frac{\rmd^2 N}{p_T\rmd p_T \rmd y}\left(1+2\sum_{n=1}^{\infty}v_n\cos\left(n(\phi-\Psi)\right) \right),
%\end{equation}
%where $E_H$ is the energy of the hadron and $\phi$ is its azimuthal emission angle relative to the reaction plane angle, $\Psi$ 
% \begin{figure}
%     \centering
%     \includegraphics[width=\linewidth]{fig/pp_PYTHIAvsALICE.pdf}
%     \caption{$p_T$-differential cross section of prompt $D_0$ mesons in pp collisions at $\sqrt{s}=5.02$ TeV in the rapidity interval $-0.5<\eta<0.5$. The result from the PYTHIA simulation(red line) is compared to that measured by the ALICE collaboration(black dots).}
%     \label{fig:pp_Dmeson}
% \end{figure}

Our results for $D$-mesons are summarized in Fig.~\ref{fig:PbPb-D0}. Fair agreement with ALICE and CMS data at mid-rapidity is found, keeping in mind that no parameter has been tuned specifically to these data. The description is better within the VCP constraints for the $T$-matrix interactions, which is reassuring given that the underlying spatial diffusion coefficient aligns better with the pertinent lQCD data. The larger interaction strength (smaller $\Ds$) at low sQGP temperatures appears to be vital to produce sufficient $v_2$ around its maximum, reiterating the important role of this observable in assessing the transport properties of the QCD medium. {Hadronization remains a key component in the description, significantly augmenting the $v_2$ and reshaping the $\raa$ (cf.~the dashed lines in the left panel of Fig.~\ref{fig:PbPb-D0}).}
Toward higher $\pT$ the $v_2$ tends to be underestimated: radiative interactions become dominant but are less effective in generating $v_2$ compared to, \eg, the purely elastic interactions in a similar framework in Ref.~\cite{He:2019vgs}. However, as pointed out in Ref.~\cite{Noronha-Hostler:2016eow}, the inclusion of event plane fluctuations, which imply a misalignment with the hydrodynamic reaction plane, may help to remedy this problem. {Thus, the combination of nonperturbative interactions at low and intermediate $\pT$ yielding to radiation of massive gluons at high $\pT$ appears to be a viable scenario (see, \eg, Refs.~\cite{Gossiaux:2010yx,Uphoff:2014hza,Grishmanovskii:2025mnc} for related discussions).}

% \begin{figure}[htbp]
%     \centering
%     \begin{subfigure}
        
%     \end{subfigure}
%     \includegraphics[scale=0.5]{fig/D0_RAA_30_50.pdf}
%     \caption{}
%     \label{fig:RAA-PbPb_RRM_HQET_30_50}
% \end{figure}

%\begin{figure}[htbp]
%\centering
%\includegraphics[width=1.\textwidth]{fig/fonll_Dmeson_Diff_VS_Radi.pdf}
%\caption{Diffusion vs. Radiation}
%\label{fig:diff-rad}
%\end{figure}

\begin{figure}[htbp]
\centering
\includegraphics[width=0.5\textwidth]{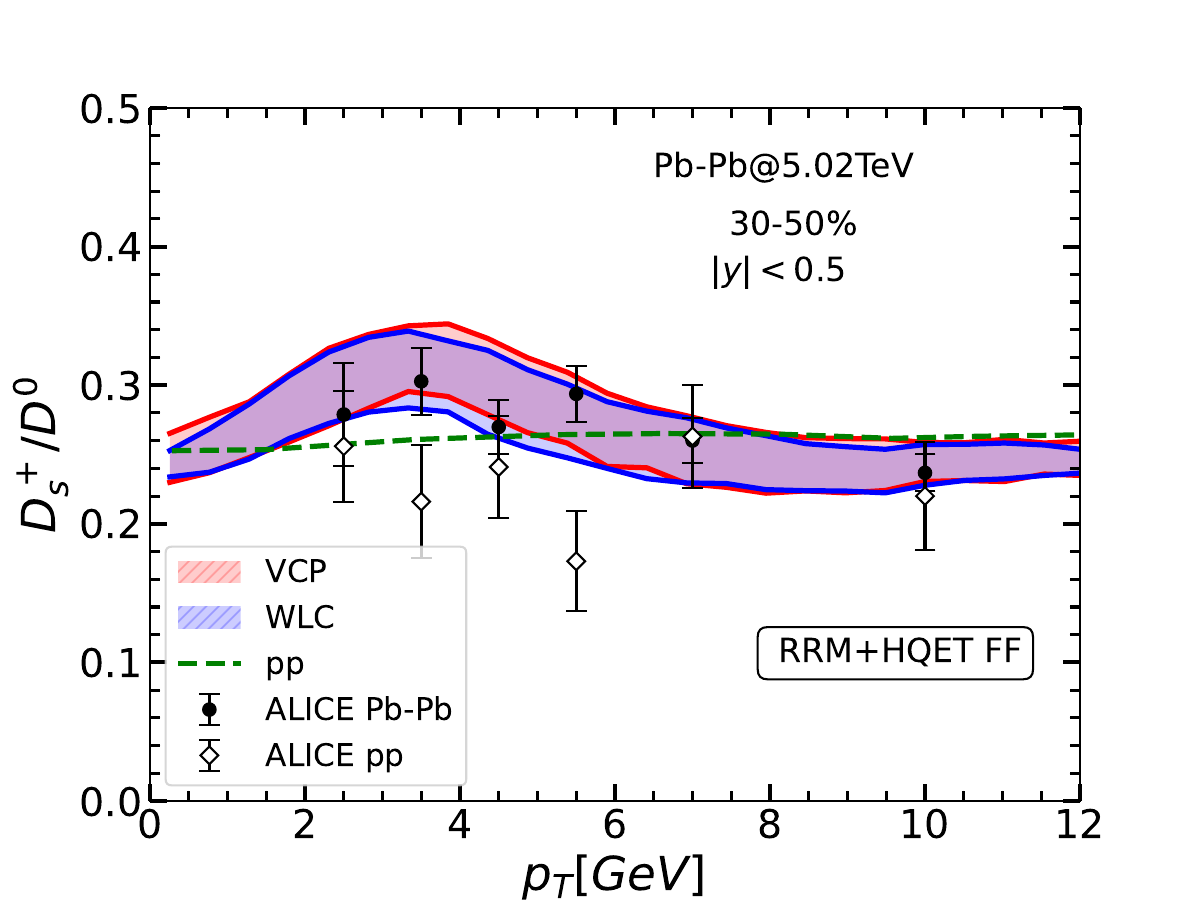}
\caption{$D_s^+/D^0$ ratio in Pb-Pb(5.02\,TeV) collisions at a hadronization temperature of $T_H$=160\,MeV. The bands reflect a range of the strange-quark mass of $m_s$=0.4-0.45\,GeV. ALICE data are taken from~\cite{ALICE:2021kfc,ALICE:2019nxm}}
\label{fig:Ds-D}
\end{figure}
%
%\begin{figure}[htbp]
%\centering
%\includegraphics[width=0.5\textwidth]{fig/BMratio_fonll_QM.pdf}
%\caption{$\Lambda_c/D_0$}
%\label{fig:B-M}
%\end{figure}
As an example of charm-hadron chemistry in Pb-Pb collisions, we display in Fig.~\ref{fig:Ds-D} our calculated $D_s^+/D^0$ ratio. An improvement over previous work~\cite{He:2019vgs} is found, mostly originating from the revised branching ratios of excited $D_s^*$ mesons into the $DK$ channel.

%%%%%%%%%%%%%%%%%%%%%%%%
{\it Conclusions}.
%%%%%%%%%%%%%%%%%%%%%%%%
We have introduced an updated framework for HF production in URHICs by merging state-of-the-art components of two existing approaches into a fully nonperturbative calculation of micro- and macro-physics. Key advances include the use of HQ transport coefficients with recent lattice constraints and a refined implementation of gluon radiation. The resulting HQ transport framework is the most comprehensive to date and represents a significant advance in the theoretical modeling of HQ dynamics in hot and dense QCD matter.  Without parameter tuning, 
%Within the same EoS with micorscopiC inteactions while also advancing i connection constraints from lattice QCD on the microphysics of the interactions to experimental data
a fair agreement with HF meson data has been found which can serve as a controlled starting point for further systematic improvements. In particular, the sensitivity to the underlying interaction strength encoded in the HQ diffusion coefficient has been highlighted. Other effects not included here, such as diffusion in pre-equilibrium and hadronic phases or event-by-event fluctuations in the hydro evolution, are presumably sub-leading but can be readily incorporated and possibly improve the description of experimental data. Work in this direction is underway.

%%%%%%%%%%%%%%%%%%%%%%%%
{\bf Acknowledgment}.
%%%%%%%%%%%%%%%%%%%%%%%%%
We thank Cameron Dean, Yen-Jie Lee, Ivan Vitev and Ramona Vogt for valuable discussions, and Zhanduo Tang for providing the $T$-matrix transport coefficients.
This work was supported by the U.S. Department of Energy through the Topical Collaboration in Nuclear Theory on {\it Heavy-Flavor Theory (HEFTY) for QCD Matter} under award no. DE-SC0023547, the DOE Office of Nuclear Physics, grant No. DE-FG02-05ER41367 (SAB), by the U.S. National Science Foundation under grant no. PHY-2209335, and by the DOE National Energy Research Scientific Computing Center (NERSC) under award NP-ERCAP0026721.

\newpage

\bibliography{HEFTY-bib}

@article{Heinz:2013th,
    author = "Heinz, Ulrich and Snellings, Raimond",
    title = "{Collective flow and viscosity in relativistic heavy-ion collisions}",
    eprint = "1301.2826",
    archivePrefix = "arXiv",
    primaryClass = "nucl-th",
    doi = "10.1146/annurev-nucl-102212-170540",
    journal = "Ann. Rev. Nucl. Part. Sci.",
    volume = "63",
    pages = "123--151",
    year = "2013"
}

@article{Shuryak:2014zxa,
    author = "Shuryak, Edward",
    title = "{Strongly coupled quark-gluon plasma in heavy ion collisions}",
    eprint = "1412.8393",
    archivePrefix = "arXiv",
    primaryClass = "hep-ph",
    doi = "10.1103/RevModPhys.89.035001",
    journal = "Rev. Mod. Phys.",
    volume = "89",
    pages = "035001",
    year = "2017"
}

@article{ParticleDataGroup:2018ovx,
    author = "Tanabashi, M. and others",
    collaboration = "Particle Data Group",
    title = "{Review of Particle Physics}",
    doi = "10.1103/PhysRevD.98.030001",
    journal = "Phys. Rev. D",
    volume = "98",
    number = "3",
    pages = "030001",
    year = "2018"
}

@article{Das:2024vac,
    author = "Das, Santosh K. and Torres-Rincon, Juan M. and Rapp, Ralf",
    title = "{Charm and bottom hadrons in hot hadronic matter}",
    eprint = "2406.13286",
    archivePrefix = "arXiv",
    primaryClass = "hep-ph",
    doi = "10.1016/j.physrep.2025.05.002",
    journal = "Phys. Rept.",
    volume = "1129-1131",
    pages = "1--53",
    year = "2025"
}

@article{PHENIX:2006iih,
    author = "Adare, A. and others",
    collaboration = "PHENIX",
    title = "{Energy Loss and Flow of Heavy Quarks in Au+Au Collisions at s(NN)**(1/2) = 200-GeV}",
    eprint = "nucl-ex/0611018",
    archivePrefix = "arXiv",
    doi = "10.1103/PhysRevLett.98.172301",
    journal = "Phys. Rev. Lett.",
    volume = "98",
    pages = "172301",
    year = "2007"
}

@article{ALICE:2020iug,
    author = "Acharya, Shreyasi and others",
    collaboration = "ALICE",
    title = "{Transverse-momentum and event-shape dependence of D-meson flow harmonics in Pb\textendash{}Pb collisions at $\sqrt {s_{NN}}$ = 5.02 TeV}",
    eprint = "2005.11131",
    archivePrefix = "arXiv",
    primaryClass = "nucl-ex",
    reportNumber = "CERN-EP-2020-082",
    doi = "10.1016/j.physletb.2020.136054",
    journal = "Phys. Lett. B",
    volume = "813",
    pages = "136054",
    year = "2021"
}

@article{ALICE:2021bib,
    author = "Acharya, Shreyasi and others",
    collaboration = "ALICE",
    title = "{Constraining hadronization mechanisms with \ensuremath{\Lambda}c+/D0 production ratios in Pb\textendash{}Pb collisions at sNN=5.02 TeV}",
    eprint = "2112.08156",
    archivePrefix = "arXiv",
    primaryClass = "nucl-ex",
    reportNumber = "CERN-EP-2021-260",
    doi = "10.1016/j.physletb.2023.137796",
    journal = "Phys. Lett. B",
    volume = "839",
    pages = "137796",
    year = "2023"
}

@article{ALICE:2021rxa,
    author = "Acharya, Shreyasi and others",
    collaboration = "ALICE",
    title = "{Prompt D$^{0}$, D$^{+}$, and D$^{*+}$ production in Pb\textendash{}Pb collisions at $ \sqrt{s_{\mathrm{NN}}} $ = 5.02 TeV}",
    eprint = "2110.09420",
    archivePrefix = "arXiv",
    primaryClass = "nucl-ex",
    reportNumber = "CERN-EP-2021-213",
    doi = "10.1007/JHEP01(2022)174",
    journal = "JHEP",
    volume = "01",
    pages = "174",
    year = "2022"
}

@article{ALICE:2021kfc,
    author = "Acharya, Shreyasi and others",
    collaboration = "ALICE",
    title = "{Measurement of prompt $D_s^+$-meson production and azimuthal anisotropy in Pb{\textendash}Pb collisions at $\sqrt {s_{NN}}$=5.02TeV}",
    eprint = "2110.10006",
    archivePrefix = "arXiv",
    primaryClass = "nucl-ex",
    reportNumber = "CERN-EP-2021-187",
    doi = "10.1016/j.physletb.2022.136986",
    journal = "Phys. Lett. B",
    volume = "827",
    pages = "136986",
    year = "2022"
}

@article{ALICE:2022exq,
    author = "Acharya, Shreyasi and others",
    collaboration = "ALICE",
    title = "{First measurement of {\ensuremath{\Lambda}}c+ production down to pT=0 in pp and p-Pb collisions at sNN=5.02 TeV}",
    eprint = "2211.14032",
    archivePrefix = "arXiv",
    primaryClass = "nucl-ex",
    reportNumber = "CERN-EP-2022-261",
    doi = "10.1103/PhysRevC.107.064901",
    journal = "Phys. Rev. C",
    volume = "107",
    number = "6",
    pages = "064901",
    year = "2023"
}

@article{Altenkort:2023oms,
    author = "Altenkort, Luis and Kaczmarek, Olaf and Larsen, Rasmus and Mukherjee, Swagato and Petreczky, Peter and Shu, Hai-Tao and Stendebach, Simon",
    collaboration = "HotQCD",
    title = "{Heavy Quark Diffusion from 2+1 Flavor Lattice QCD with 320~MeV Pion Mass}",
    eprint = "2302.08501",
    archivePrefix = "arXiv",
    primaryClass = "hep-lat",
    doi = "10.1103/PhysRevLett.130.231902",
    journal = "Phys. Rev. Lett.",
    volume = "130",
    number = "23",
    pages = "231902",
    year = "2023"
}

@article{Altenkort:2023eav,
    author = "Altenkort, Luis and de la Cruz, David and Kaczmarek, Olaf and Larsen, Rasmus and Moore, Guy D. and Mukherjee, Swagato and Petreczky, Peter and Shu, Hai-Tao and Stendebach, Simon",
    collaboration = "HotQCD",
    title = "{Quark Mass Dependence of Heavy Quark Diffusion Coefficient from Lattice QCD}",
    eprint = "2311.01525",
    archivePrefix = "arXiv",
    primaryClass = "hep-lat",
    doi = "10.1103/PhysRevLett.132.051902",
    journal = "Phys. Rev. Lett.",
    volume = "132",
    number = "5",
    pages = "051902",
    year = "2024"
}

@article{HotQCD:2025fbd,
    author = "Bollweg, Dennis and Dasilva Gol{\'a}n, Jorge Luis and Kaczmarek, Olaf and Larsen, Rasmus Norman and Moore, Guy D. and Mukherjee, Swagato and Petreczky, Peter and Shu, Hai-Tao and Stendebach, Simon and Weber, Johannes Heinrich",
    collaboration = "HotQCD",
    title = "{Temperature Dependence of Heavy Quark Diffusion from (2+1)-flavor Lattice QCD}",
    eprint = "2506.11958",
    archivePrefix = "arXiv",
    primaryClass = "hep-lat",
    month = "6",
    year = "2025"
}

@article{Ke:2018tsh,
    author = "Ke, Weiyao and Xu, Yingru and Bass, Steffen A.",
    title = "{Linearized Boltzmann-Langevin model for heavy quark transport in hot and dense QCD matter}",
    eprint = "1806.08848",
    archivePrefix = "arXiv",
    primaryClass = "nucl-th",
    doi = "10.1103/PhysRevC.98.064901",
    journal = "Phys. Rev. C",
    volume = "98",
    number = "6",
    pages = "064901",
    year = "2018"
}

@article{Ke:2018jem,
    author = "Ke, Weiyao and Xu, Yingru and Bass, Steffen A.",
    title = "{Modified Boltzmann approach for modeling the splitting vertices induced by the hot QCD medium in the deep Landau-Pomeranchuk-Migdal region}",
    eprint = "1810.08177",
    archivePrefix = "arXiv",
    primaryClass = "nucl-th",
    doi = "10.1103/PhysRevC.100.064911",
    journal = "Phys. Rev. C",
    volume = "100",
    number = "6",
    pages = "064911",
    year = "2019"
}

@article{Kang:2016ofv,
    author = "Kang, Zhong-Bo and Ringer, Felix and Vitev, Ivan",
    title = "{Effective field theory approach to open heavy flavor production in heavy-ion collisions}",
    eprint = "1610.02043",
    archivePrefix = "arXiv",
    primaryClass = "hep-ph",
    doi = "10.1007/JHEP03(2017)146",
    journal = "JHEP",
    volume = "03",
    pages = "146",
    year = "2017"
}

@article{Ravagli:2007xx,
    author = "Ravagli, L. and Rapp, R.",
    title = "{Quark Coalescence based on a Transport Equation}",
    eprint = "0705.0021",
    archivePrefix = "arXiv",
    primaryClass = "hep-ph",
    doi = "10.1016/j.physletb.2007.07.043",
    journal = "Phys. Lett. B",
    volume = "655",
    pages = "126--131",
    year = "2007"
}

@article{Ravagli:2008rt,
    author = "Ravagli, L. and van Hees, H. and Rapp, R.",
    title = "{Resonance Recombination Model: A Dynamical Framework for Hadronization}",
    eprint = "0806.2055",
    archivePrefix = "arXiv",
    primaryClass = "hep-ph",
    doi = "10.1103/PhysRevC.79.064902",
    journal = "Phys. Rev. C",
    volume = "79",
    pages = "064902",
    year = "2009"
}

@article{He:2011qa,
    author = "He, Min and Fries, Rainer J. and Rapp, Ralf",
    title = "{Heavy-Quark Diffusion and Hadronization in Quark-Gluon Plasma}",
    eprint = "1106.6006",
    archivePrefix = "arXiv",
    primaryClass = "nucl-th",
    doi = "10.1103/PhysRevC.86.014903",
    journal = "Phys. Rev. C",
    volume = "86",
    pages = "014903",
    year = "2012"
}

@article{He:2019tik,
    author = "He, Min and Rapp, Ralf",
    title = "{Charm-Baryon Production in Proton-Proton Collisions}",
    eprint = "1902.08889",
    archivePrefix = "arXiv",
    primaryClass = "nucl-th",
    doi = "10.1016/j.physletb.2019.06.004",
    journal = "Phys. Lett. B",
    volume = "795",
    pages = "117--121",
    year = "2019"
}

@article{He:2019vgs,
    author = "He, Min and Rapp, Ralf",
    title = "{Hadronization and Charm-Hadron Ratios in Heavy-Ion Collisions}",
    eprint = "1905.09216",
    archivePrefix = "arXiv",
    primaryClass = "nucl-th",
    doi = "10.1103/PhysRevLett.124.042301",
    journal = "Phys. Rev. Lett.",
    volume = "124",
    number = "4",
    pages = "042301",
    year = "2020"
}

@article{Rapp:2018qla,
    author = "Beraudo, A. and others",
    editor = "Rapp, R. and Gossiaux, P. B. and Andronic, A. and Averbeck, R. and Masciocchi, S.",
    title = "{Extraction of Heavy-Flavor Transport Coefficients in QCD Matter}",
    eprint = "1803.03824",
    archivePrefix = "arXiv",
    primaryClass = "nucl-th",
    doi = "10.1016/j.nuclphysa.2018.09.002",
    journal = "Nucl. Phys. A",
    volume = "979",
    pages = "21--86",
    year = "2018"
}

@article{Zhao:2023nrz,
    author = "Zhao, Jiaxing and others",
    title = "{Hadronization of heavy quarks}",
    eprint = "2311.10621",
    archivePrefix = "arXiv",
    primaryClass = "hep-ph",
    doi = "10.1103/PhysRevC.109.054912",
    journal = "Phys. Rev. C",
    volume = "109",
    number = "5",
    pages = "054912",
    year = "2024"
}

@article{He:2022ywp,
    author = "He, Min and van Hees, Hendrik and Rapp, Ralf",
    title = "{Heavy-quark diffusion in the quark{\textendash}gluon plasma}",
    eprint = "2204.09299",
    archivePrefix = "arXiv",
    primaryClass = "hep-ph",
    doi = "10.1016/j.ppnp.2023.104020",
    journal = "Prog. Part. Nucl. Phys.",
    volume = "130",
    pages = "104020",
    year = "2023"
}

@article{ALICE:2019nxm,
    author = "Acharya, Shreyasi and others",
    collaboration = "ALICE",
    title = "{Measurement of ${{\mathrm{D}}^0}$ , ${{\mathrm{D}}^+}$ , ${{\mathrm{D}}^{*+}}$ and ${{\mathrm{D}}^+_{\mathrm{s}}}$ production in pp collisions at ${\sqrt{{\textit{s}}}~=~5.02~{\text {TeV}}}$ with ALICE}",
    eprint = "1901.07979",
    archivePrefix = "arXiv",
    primaryClass = "nucl-ex",
    reportNumber = "CERN-EP-2019-004",
    doi = "10.1140/epjc/s10052-019-6873-6",
    journal = "Eur. Phys. J. C",
    volume = "79",
    number = "5",
    pages = "388",
    year = "2019"
}

@article{CMS:2017vhp,
    author = "Sirunyan, Albert M and others",
    collaboration = "CMS",
    title = "{Measurement of prompt $D^0$ meson azimuthal anisotropy in Pb-Pb collisions at $\sqrt{{s}_{NN}}$ = 5.02 TeV}",
    eprint = "1708.03497",
    archivePrefix = "arXiv",
    primaryClass = "nucl-ex",
    reportNumber = "CMS-HIN-16-007, CERN-EP-2017-174",
    doi = "10.1103/PhysRevLett.120.202301",
    journal = "Phys. Rev. Lett.",
    volume = "120",
    number = "20",
    pages = "202301",
    year = "2018"
}

@article{CMS:2020bnz,
    author = "Sirunyan, Albert M and others",
    collaboration = "CMS",
    title = "{Measurement of prompt ${\mathrm{D^0}}$ and ${\mathrm{\overline{D}}{}^0}$ meson azimuthal anisotropy and search for strong electric fields in PbPb collisions at $\sqrt{s_\mathrm{NN}} =$  5.02 TeV}",
    eprint = "2009.12628",
    archivePrefix = "arXiv",
    primaryClass = "hep-ex",
    reportNumber = "CMS-HIN-19-008, CERN-EP-2020-155",
    doi = "10.1016/j.physletb.2021.136253",
    journal = "Phys. Lett. B",
    volume = "816",
    pages = "136253",
    year = "2021"
}

@article{Tang:2023tkm,
    author = "Tang, Zhanduo and Mukherjee, Swagato and Petreczky, Peter and Rapp, Ralf",
    title = "{T-matrix analysis of static Wilson line correlators from lattice QCD at finite temperature}",
    eprint = "2310.18864",
    archivePrefix = "arXiv",
    primaryClass = "hep-lat",
    doi = "10.1140/epja/s10050-024-01310-w",
    journal = "Eur. Phys. J. A",
    volume = "60",
    number = "4",
    pages = "92",
    year = "2024"
}

@article{Liu:2018syc,
    author = "Liu, Shuai Y. F. and He, Min and Rapp, Ralf",
    title = "{Probing the in-Medium QCD Force by Open Heavy-Flavor Observables}",
    eprint = "1806.05669",
    archivePrefix = "arXiv",
    primaryClass = "nucl-th",
    doi = "10.1103/PhysRevC.99.055201",
    journal = "Phys. Rev. C",
    volume = "99",
    number = "5",
    pages = "055201",
    year = "2019"
}

@article{Liu:2020dlt,
    author = "Liu, Shuai Y. F. and Rapp, Ralf",
    title = "{Nonperturbative Effects on Radiative Energy Loss of Heavy Quarks}",
    eprint = "2003.12536",
    archivePrefix = "arXiv",
    primaryClass = "nucl-th",
    doi = "10.1007/JHEP08(2020)168",
    journal = "JHEP",
    volume = "08",
    pages = "168",
    year = "2020"
}

@article{Liu:2017qah,
    author = "Liu, Shuai Y. F. and Rapp, Ralf",
    title = "{$T$-matrix Approach to Quark-Gluon Plasma}",
    eprint = "1711.03282",
    archivePrefix = "arXiv",
    primaryClass = "nucl-th",
    doi = "10.1103/PhysRevC.97.034918",
    journal = "Phys. Rev. C",
    volume = "97",
    number = "3",
    pages = "034918",
    year = "2018"
}

@article{Luttinger:1960ua,
    author = "Luttinger, J. M. and Ward, John Clive",
    title = "{Ground state energy of a many fermion system. 2.}",
    doi = "10.1103/PhysRev.118.1417",
    journal = "Phys. Rev.",
    volume = "118",
    pages = "1417--1427",
    year = "1960"
}

@article{Baym:1961zz,
    author = "Baym, Gordon and Kadanoff, Leo P.",
    title = "{Conservation Laws and Correlation Functions}",
    doi = "10.1103/PhysRev.124.287",
    journal = "Phys. Rev.",
    volume = "124",
    pages = "287--299",
    year = "1961"
}

@article{Ke:2020clc,
    author = "Ke, Weiyao and Wang, Xin-Nian",
    title = "{QGP modification to single inclusive jets in a calibrated transport model}",
    eprint = "2010.13680",
    archivePrefix = "arXiv",
    primaryClass = "hep-ph",
    doi = "10.1007/JHEP05(2021)041",
    journal = "JHEP",
    volume = "05",
    pages = "041",
    year = "2021"
}

@article{Arnold:2002ja,
    author = "Arnold, Peter Brockway and Moore, Guy D. and Yaffe, Laurence G.",
    title = "{Photon and gluon emission in relativistic plasmas}",
    eprint = "hep-ph/0204343",
    archivePrefix = "arXiv",
    reportNumber = "UW-PT-02-06",
    doi = "10.1088/1126-6708/2002/06/030",
    journal = "JHEP",
    volume = "06",
    pages = "030",
    year = "2002"
}

@article{Arnold:2002zm,
    author = "Arnold, Peter Brockway and Moore, Guy D. and Yaffe, Laurence G.",
    title = "{Effective kinetic theory for high temperature gauge theories}",
    eprint = "hep-ph/0209353",
    archivePrefix = "arXiv",
    doi = "10.1088/1126-6708/2003/01/030",
    journal = "JHEP",
    volume = "01",
    pages = "030",
    year = "2003"
}

@article{Moreland:2014oya,
    author = "Moreland, J. Scott and Bernhard, Jonah E. and Bass, Steffen A.",
    title = "{Alternative ansatz to wounded nucleon and binary collision scaling in high-energy nuclear collisions}",
    eprint = "1412.4708",
    archivePrefix = "arXiv",
    primaryClass = "nucl-th",
    doi = "10.1103/PhysRevC.92.011901",
    journal = "Phys. Rev. C",
    volume = "92",
    number = "1",
    pages = "011901",
    year = "2015"
}

@article{Liu:2015nwa,
    author = "Liu, Jia and Shen, Chun and Heinz, Ulrich",
    title = "{Pre-equilibrium evolution effects on heavy-ion collision observables}",
    eprint = "1504.02160",
    archivePrefix = "arXiv",
    primaryClass = "nucl-th",
    doi = "10.1103/PhysRevC.91.064906",
    journal = "Phys. Rev. C",
    volume = "91",
    number = "6",
    pages = "064906",
    year = "2015",
    note = "[Erratum: Phys.Rev.C 92, 049904 (2015)]"
}

@article{Shen:2014vra,
    author = "Shen, Chun and Qiu, Zhi and Song, Huichao and Bernhard, Jonah and Bass, Steffen and Heinz, Ulrich",
    title = "{The iEBE-VISHNU code package for relativistic heavy-ion collisions}",
    eprint = "1409.8164",
    archivePrefix = "arXiv",
    primaryClass = "nucl-th",
    doi = "10.1016/j.cpc.2015.08.039",
    journal = "Comput. Phys. Commun.",
    volume = "199",
    pages = "61--85",
    year = "2016"
}

@article{HotQCD:2014kol,
    author = "Bazavov, A. and others",
    collaboration = "HotQCD",
    title = "{Equation of state in ( 2+1 )-flavor QCD}",
    eprint = "1407.6387",
    archivePrefix = "arXiv",
    primaryClass = "hep-lat",
    reportNumber = "BNL-105928-2014-JA",
    doi = "10.1103/PhysRevD.90.094503",
    journal = "Phys. Rev. D",
    volume = "90",
    pages = "094503",
    year = "2014"
}

@article{Bass:1998ca,
    author = "Bass, S. A. and others",
    title = "{Microscopic models for ultrarelativistic heavy ion collisions}",
    eprint = "nucl-th/9803035",
    archivePrefix = "arXiv",
    doi = "10.1016/S0146-6410(98)00058-1",
    journal = "Prog. Part. Nucl. Phys.",
    volume = "41",
    pages = "255--369",
    year = "1998"
}

@article{Bleicher:1999xi,
    author = "Bleicher, M. and others",
    title = "{Relativistic hadron hadron collisions in the ultrarelativistic quantum molecular dynamics model}",
    eprint = "hep-ph/9909407",
    archivePrefix = "arXiv",
    doi = "10.1088/0954-3899/25/9/308",
    journal = "J. Phys. G",
    volume = "25",
    pages = "1859--1896",
    year = "1999"
}

@article{Bernhard:2019bmu,
    author = "Bernhard, Jonah E. and Moreland, J. Scott and Bass, Steffen A.",
    title = "{Bayesian estimation of the specific shear and bulk viscosity of quark{\textendash}gluon plasma}",
    doi = "10.1038/s41567-019-0611-8",
    journal = "Nature Phys.",
    volume = "15",
    number = "11",
    pages = "1113--1117",
    year = "2019"
}

@article{Noronha-Hostler:2016eow,
    author = "Noronha-Hostler, Jacquelyn and Betz, Barbara and Noronha, Jorge and Gyulassy, Miklos",
    title = "{Event-by-event hydrodynamics $+$ jet energy loss: A solution to the $R_{AA} \otimes v_2$ puzzle}",
    eprint = "1602.03788",
    archivePrefix = "arXiv",
    primaryClass = "nucl-th",
    doi = "10.1103/PhysRevLett.116.252301",
    journal = "Phys. Rev. Lett.",
    volume = "116",
    number = "25",
    pages = "252301",
    year = "2016"
}

@article{Chesler:2013cqa,
    author = "Chesler, Paul M. and Lekaveckas, Mindaugas and Rajagopal, Krishna",
    title = "{Heavy quark energy loss far from equilibrium in a strongly coupled collision}",
    eprint = "1306.0564",
    archivePrefix = "arXiv",
    primaryClass = "hep-ph",
    reportNumber = "CERN-PH-TH-2013-109",
    doi = "10.1007/JHEP10(2013)013",
    journal = "JHEP",
    volume = "10",
    pages = "013",
    year = "2013"
}

@article{Mrowczynski:2017kso,
    author = "Mrowczynski, Stanislaw",
    title = "{Heavy Quarks in Turbulent QCD Plasmas}",
    eprint = "1706.03127",
    archivePrefix = "arXiv",
    primaryClass = "nucl-th",
    doi = "10.1140/epja/i2018-12478-5",
    journal = "Eur. Phys. J. A",
    volume = "54",
    number = "3",
    pages = "43",
    year = "2018"
}

@article{Ebert:2011kk,
    author = "Ebert, D. and Faustov, R. N. and Galkin, V. O.",
    title = "{Spectroscopy and Regge trajectories of heavy baryons in the relativistic quark-diquark picture}",
    eprint = "1105.0583",
    archivePrefix = "arXiv",
    primaryClass = "hep-ph",
    reportNumber = "HU-EP-11-21",
    doi = "10.1103/PhysRevD.84.014025",
    journal = "Phys. Rev. D",
    volume = "84",
    pages = "014025",
    year = "2011"
}

@article{Fu:2025,
    author = "Fu, Y. and Krishna, T. and Ke, W. and Bass, S. and Rapp, R.",
    title = "{Heavy-Flavor Tranport Approach for Ultararelativistic Heavy-Ion Collisions}" ,     
     year = "in preparation (2025)"
}

@article{Cacciari:2001td,
    author = "Cacciari, Matteo and Frixione, Stefano and Nason, Paolo",
    title = "{The p(T) spectrum in heavy flavor photoproduction}",
    eprint = "hep-ph/0102134",
    archivePrefix = "arXiv",
    reportNumber = "BICOCCA-FT-01-01, GEF-TH-2-01, YITP-SB-01-01",
    doi = "10.1088/1126-6708/2001/03/006",
    journal = "JHEP",
    volume = "03",
    pages = "006",
    year = "2001"
}

@article{Danielewicz:1984ww,
    author = "Danielewicz, P. and Gyulassy, M.",
    title = "{Dissipative Phenomena in Quark Gluon Plasmas}",
    doi = "10.1103/PhysRevD.31.53",
    journal = "Phys. Rev. D",
    volume = "31",
    pages = "53--62",
    year = "1985"
}

@article{Kovtun:2004de,
    author = "Kovtun, P. and Son, Dan T. and Starinets, Andrei O.",
    title = "{Viscosity in strongly interacting quantum field theories from black hole physics}",
    eprint = "hep-th/0405231",
    archivePrefix = "arXiv",
    reportNumber = "INT-PUB-04-09, UW-PT-04-04",
    doi = "10.1103/PhysRevLett.94.111601",
    journal = "Phys. Rev. Lett.",
    volume = "94",
    pages = "111601",
    year = "2005"
}

@article{Emelyanov:1997guf,
    author = "Emel'yanov, V. and Khodinov, A. and Klein, S. R. and Vogt, R.",
    title = "{Charm quark production in noncentral heavy ion collisions}",
    eprint = "nucl-th/9706085",
    archivePrefix = "arXiv",
    reportNumber = "LBL-40398, LBNL-40398",
    doi = "10.1103/PhysRevC.56.2726",
    journal = "Phys. Rev. C",
    volume = "56",
    pages = "2726--2735",
    year = "1997"
}

@article{Gossiaux:2010yx,
    author = "Gossiaux, P. B. and Aichelin, J. and Gousset, T. and Guiho, V.",
    editor = "Fraga, Eduardo and Kodama, Takeshi and Padula, Sandra and Takahashi, Jun",
    title = "{Competition of Heavy Quark Radiative and Collisional Energy Loss in Deconfined Matter}",
    eprint = "1001.4166",
    archivePrefix = "arXiv",
    primaryClass = "hep-ph",
    doi = "10.1088/0954-3899/37/9/094019",
    journal = "J. Phys. G",
    volume = "37",
    pages = "094019",
    year = "2010"
}

@article{Uphoff:2014hza,
    author = "Uphoff, Jan and Fochler, Oliver and Xu, Zhe and Greiner, Carsten",
    title = "{Elastic and radiative heavy quark interactions in ultra-relativistic heavy-ion collisions}",
    eprint = "1408.2964",
    archivePrefix = "arXiv",
    primaryClass = "hep-ph",
    doi = "10.1088/0954-3899/42/11/115106",
    journal = "J. Phys. G",
    volume = "42",
    number = "11",
    pages = "115106",
    year = "2015"
}

@article{Grishmanovskii:2025mnc,
    author = "Grishmanovskii, Ilia and Song, Taesoo and Greiner, Carsten and Bratkovskaya, Elena",
    title = "{Transport coefficients of heavy quarks by elastic and radiative scatterings in the strongly interacting quark-gluon plasma}",
    eprint = "2503.22311",
    archivePrefix = "arXiv",
    primaryClass = "hep-ph",
    doi = "10.1103/rb24-dg58",
    journal = "Phys. Rev. D",
    volume = "112",
    number = "1",
    pages = "014042",
    year = "2025"
}

@article{CMS:2017qjw,
    author = "Sirunyan, Albert M and others",
    collaboration = "CMS",
    title = "{Nuclear modification factor of D$^0$ mesons in PbPb collisions at  $\sqrt{s_\mathrm{NN}} = 5.02$ TeV}",
    eprint = "1708.04962",
    archivePrefix = "arXiv",
    primaryClass = "nucl-ex",
    reportNumber = "CMS-HIN-16-001, CERN-EP-2017-186",
    doi = "10.1016/j.physletb.2018.05.074",
    journal = "Phys. Lett. B",
    volume = "782",
    pages = "474--496",
    year = "2018"
}

\end{document}